\documentclass[10pt]{article}
 \usepackage{authblk}
 \usepackage{graphicx}
 \usepackage{caption}
 \usepackage{subfigure}
 \usepackage{amsfonts}
 \usepackage[top=0.75in, bottom=0.75in, left=1.0in, right=1.0in]{geometry}

\newcommand{\expect}{\mathbb{E}}
 
\begin{document}
\pagenumbering{gobble}
 
\title
{
Spectral Anomaly Detection in Very Large Graphs\\%
{\Large Models, Noise, and Computational Complexity}%
} 
\author
{
	Benjamin A. Miller, Nicholas Arcolano\thanks{Currently at FitnessKeeper, Inc.: \texttt{nicholas.arcolano@runkeeper.com}}, Michael M. Wolf\thanks{Currently at Sandia National Laboratories: \texttt{mmwolf@sandia.gov}}, and Nadya T. Bliss\thanks{Currently at Arizona State University: \texttt{nadya.bliss@asu.edu}}\\
        MIT Lincoln Laboratory\\
        \texttt{bamiller@ll.mit.edu}
}   
\date{\vspace{-5ex}}

\maketitle

Data analysis tasks considering entities and their connections and interactions are inherent in numerous application domains. For example, computer security applications may consider communication between machines to determine abnormal behavior in the network \cite{eigenspaceAnomalyDetection}, and medical imaging may analyze the functional and anatomical connectivity between brain regions to determine the regions affected by a neurological condition \cite{Archana2013}. These tasks are naturally formulated as graph analysis problems, and in applications as varied as biology, sociology, and network security, one technical problem of interest is the detection of a small subregion of the graph in which the connections are significantly different than expected under normal conditions.
 
Signal processing for graphs (SPG) is a recent technical effort to address this problem in an application-agnostic setting \cite{SPGLLJ}. SPG provides a statistical framework addressing the subgraph detection problem in the context of traditional detection theory, where the objective is to resolve a binary hypothesis test. Under the null hypothesis, the observed data---i.e., an observed graph---is drawn from a ``noise'' distribution representing typical activity, whereas under the alternative hypothesis the graph also contains a subset of vertices whose connectivity significantly deviates from the expectation. Inspired by modularity analysis \cite{Newman06}, the framework is based on analysis of graph residuals---the difference between the observed and expected topology of the graph---and is designed in a modular way to enable a variety of models and techniques. The processing chain includes (1) fitting (possibly dynamic) graph data to an expected connectivity model, (2) computing the graph residuals (possibly aggregating over time), (3) projection into a low-dimensional space, (4) computation of a detection statistic to determine the presence of an anomaly, and (5) identification of the vertices (entities) exhibiting the anomalous behavior. Specific attention is paid to spectral methods \cite{Miller2014}, as they provide natural metrics for signal and noise power. Thus, the methods used here focus on computation and analysis of the principal eigenvectors of $A-\expect\left[A\right]$, where $A$ is the adjacency matrix of the graph.

A recent series of studies aimed to expand the basis of the framework in three technical dimensions. One area of interest was the development of techniques that incorporate graph dynamics and attributes into the residuals analysis, in a way that scales well to very large graphs. As graphs of interest are often obtained through noisy or unreliable sources, another important issue is the impact of uncertainty and corruption on subgraph detection performance. Finally, since many application areas consider graphs that will not fit into memory on a desktop computer, understanding how high-performance computing technologies can best assist large-scale graph residuals analysis is an important aspect of this work. This abstract provides a brief summary outlining results with respect to each of these technical challenges.

To integrate attributes into the residuals analysis framework, we use a generalized linear model for edge probabilities, based on the vertex metadata \cite{ISI2013}. We formulate the model of the probability of an edge occurring from vertex $i$ to vertex $j$ as
\begin{equation}
p_{ij}=g\left(x_i^T\beta_1+x_j^T\beta_2+x_{ij}^T\beta_3\right),
\end{equation}
where $g$ is a link function, $x_i$ and $x_j$ represent attribute vectors for the vertices, $x_{ij}$ is a feature vector for the pair, and the $\beta$ vectors appropriately weight the attributes and are learned from the data. For $g$, we use the logistic function $g(x)=\left(1+\exp\left(-x\right)\right)^{-1}$. This presents a problem for spectral analysis when graphs become very large: $A$ will typically be sparse, but $\expect\left[A\right]$ will be dense. In a restricted setting and assuming that probabilities are low, however, we can leverage a structure that will allow fast matrix-vector multiplications, thus enabling efficient computation of the principal residuals space. First, if we approximate the logistic function as an exponential function (a reasonable approximation for small probabilities), the edge probability becomes the product of three terms: one based entirely on the source vertex, one on the destination vertex, and one on the properties of the pair. If we also restrict the vertex-pair attributes to pairs of vertex categories (e.g., a math paper cites a physics paper), then the probability matrix $\expect\left[A\right]$ will have a low-rank structure that can be exploited to achieve fast matrix-vector multiplications. This will have the same structure as the degree-corrected stochastic blockmodel \cite{Zhang2014}. This model also yields an efficient parameter estimation procedure based on moment matching. 

\begin{figure}
        \centering
\includegraphics[width=2.1in]{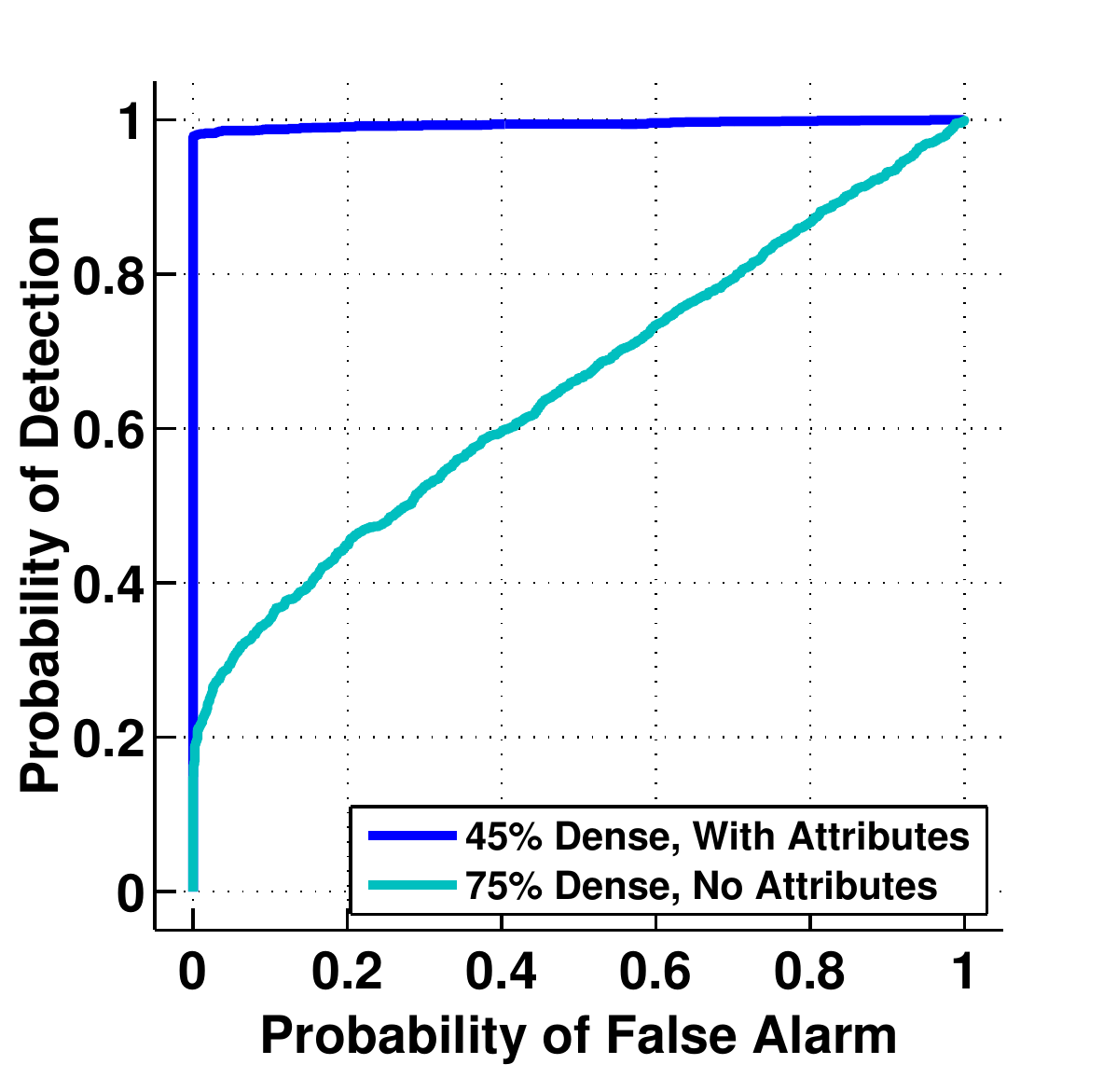}
\includegraphics[width=2.0in]{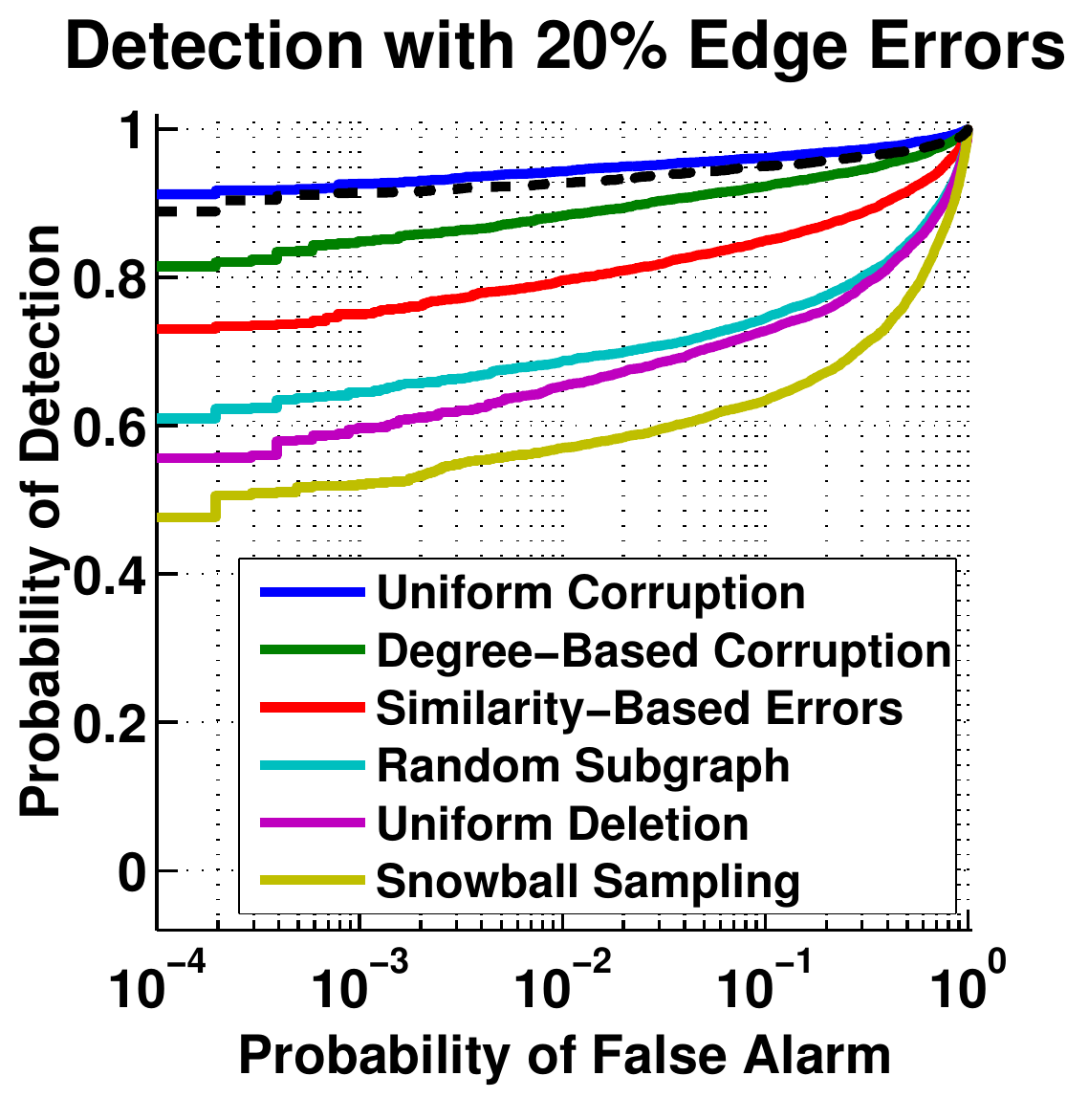}
\includegraphics[width=2.0in]{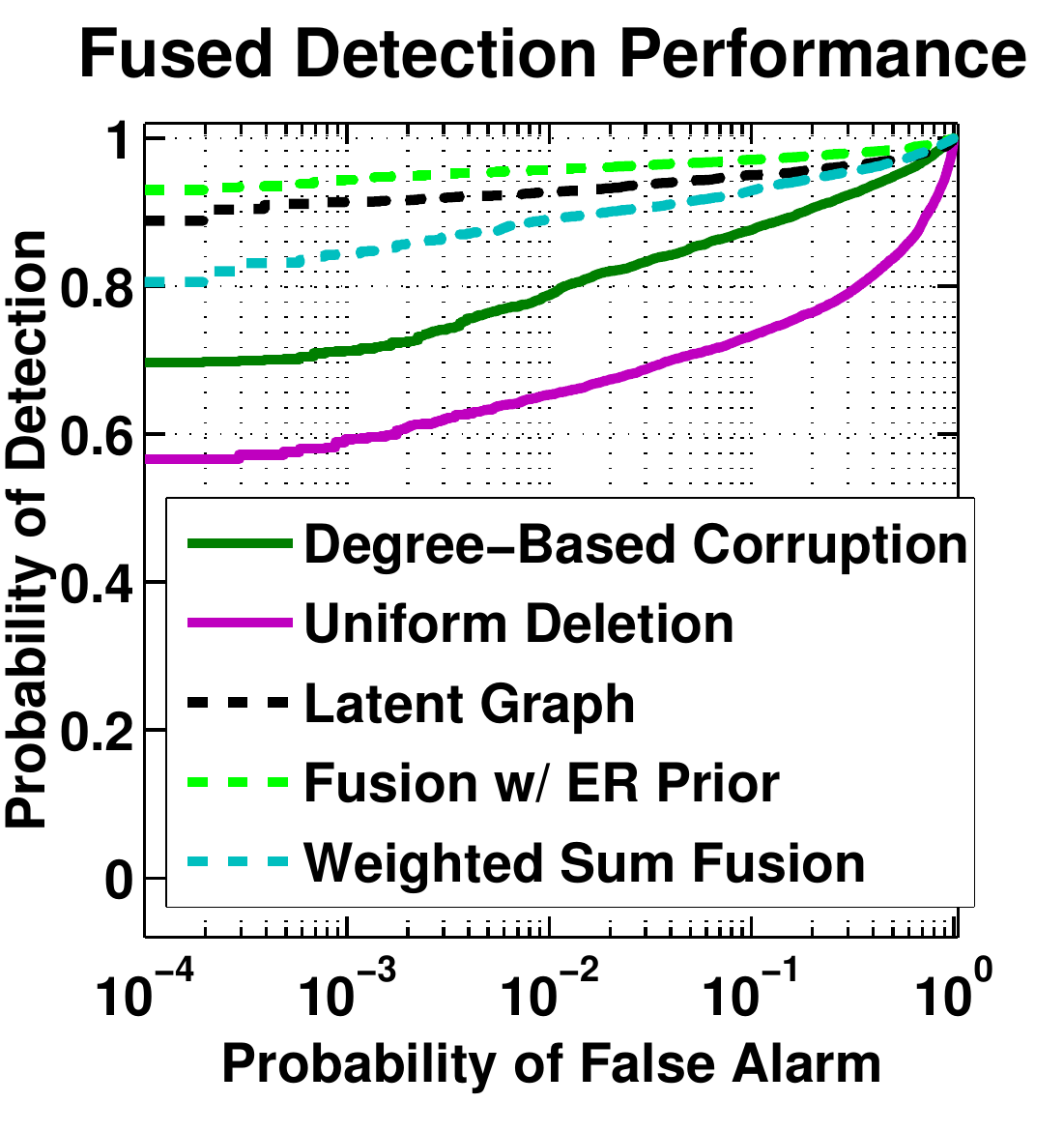}\ \ \ 
        \caption{Detection performance with attributes and uncertainty. (left) Subgraph detection performance is markedly improved when attributes are considered in the expected value model. (center) Different error mechanisms yield substantially different results with respect to operating on the true graph (black dashed line). (right) Performance can be recovered using either a Bayesian method or by weighting the measurements from different uncertainty mechanisms.}
        \label{fig:analysis}
\end{figure}

For models of uncertainty and corruption, we considered recent experience with real datasets as well as those from the open literature \cite{ModelingWithMissingData}. To model issues such as sensor dropouts, we remove edges uniformly at random. To model sensor noise, we corrupt the graph with random edge addition and removal. In this situation, we consider each pair of edges and, with some probability, either remove the edge if it is present or add it if it is absent. The probability may either be uniform over the graph or based on the degrees of the vertices. Since many graphs are samplings of a population, we consider two vertex-sampling methods: selecting a random subset of vertices and the edges within that subset, and a ``snowball'' sampling method in which a random set of seed vertices is chosen and the observed graph is determined by following the links of the currently observed vertices with some probability. With attributed graphs, nodes with similar attributes can sometimes be mistaken for one another, so we consider a model in which each vertex has a feature vector, and the distance between two vertices' vectors determines the probability that a given edge connects to one when the intention is to connect to the other.

The impact of data attributes and data corruption are demonstrated in the plots in Figure \ref{fig:analysis}. On the left, a 15-vertex subgraph that densifies and disperses over time is possibly embedded into a background with a skewed degree distribution and 3 categories for each vertex, and the objective is to determine whether the embedding occurred. The background is drawn independently at random using the same model parameters at each of 8 time samples. Full experimental details are provided in \cite{ISI2013}. As demonstrated by the receiver operating characteristic (ROC) curves, when the subgraph reaches 75\% density before dispersing, but attributes are not considered, performance is not much better than chance. However, if the vertex attributes are taken into account, we achieve near-perfect performance even when the subgraph is much smaller. To quantify the impact of uncertainty and corruption, each of the uncertainty mechanisms defined previously were applied to the specific problem of detecting a 12-vertex, 85\%-dense subgraph in a 1024-vertex R-MAT background \cite{RMAT} with average degree of about 10. Normalizing each method so that the edge errors (i.e., number of missing edges plus number of incorrect edges, divided by the number of edges in the true graph) is 20\%, the impact of these mechanisms is provided in the center plot. A detailed analysis of 5 of the 6 mechanisms (all except snowball sampling) is given in \cite{ICASSP2014}. Uniform corruption actually slightly improves performance over using the true graph, since it makes the background more similar to the assumed model. Degree-based corruption adds noise that is correlated with the random background fluctuations, thus reducing performance. Similarity-based errors, as implemented here where all vertices are given a random 3-dimensional vector in the unit hypercube, has a slight ``whitening'' effect that degrades performance less than in the case of random vertex and edge removal. Snowball sampling yields the worst subgraph detection performance, as it is highly biased by the seed vertices. However, as shown in the righthand plot, performance can be recovered using multiple corrupt sources. In this example, edge deletion and degree-based corruption are fused to improve performance. In a situation where there is knowledge of the uncertainty mechanism, a Bayesian approach can be used to actually achieve better performance than on the latent graph, since the algorithm is given knowledge of the expected number of edges. If this information is not available, using a weighted sum of the individual adjacency matrices still provides a substantial increase in detection power, improving the false alarm probability at an 80\% detection rate by about 2 orders of magnitude.

\begin{figure}
    \centering        
    \includegraphics[width=3.0in]{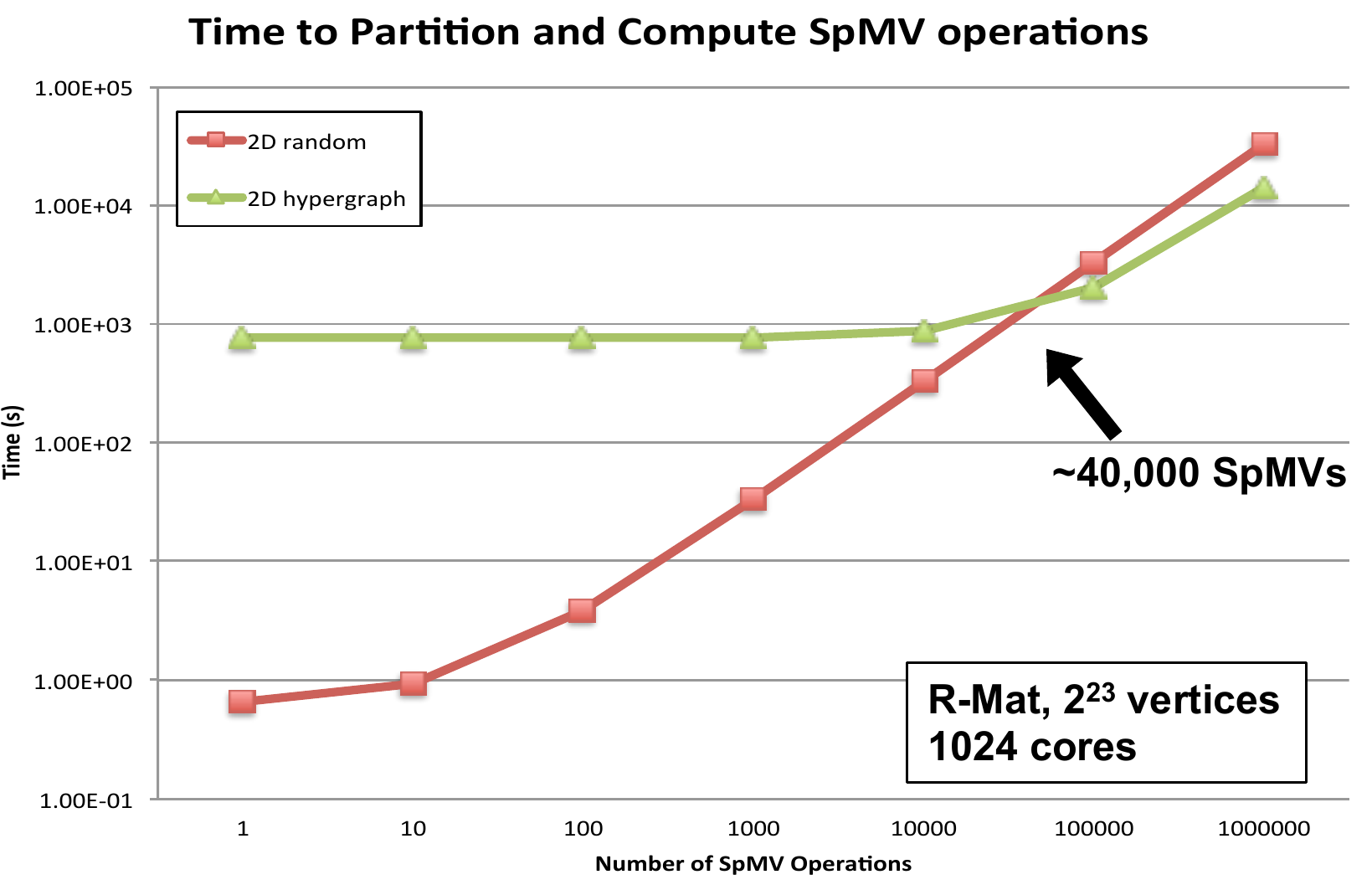}\ \ 
    \includegraphics[width=3.0in]{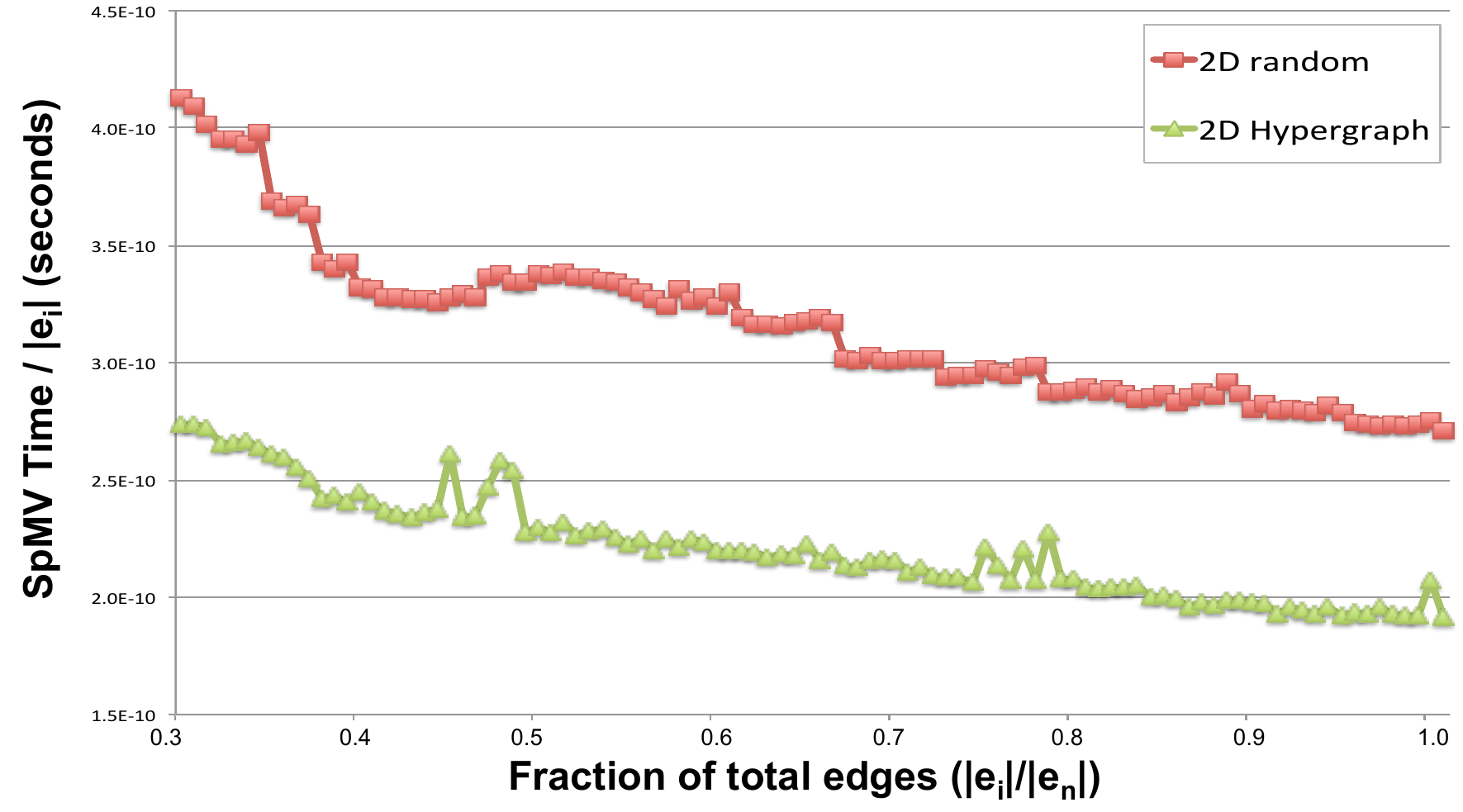}
        \caption{Impact of partitioning with reduced data. (left) In order to make hypergraph partitioning worth its computational cost, about 40,000 matrix-vector multiplications must be performed. (right) Even when partitioning is done with only 30\% of edges are available, the hypergraph partitioning method provides a significant speedup over random partitioning.}
        \label{fig:partitioning}
\end{figure}

To apply these methods to large-scale graphs, we require the ability to efficiently compute the eigenvectors of residuals matrices for large-scale graphs, which necessitates methods for appropriately dividing the data among many processes. Recent methods for 2D partitioning have shown promise for large graphs with skewed degree distributions \cite{yoo_short}, and even randomly partitioning in such a way allows computation of the top eigenvectors of multi-billion-vertex graphs in minutes \cite{HPEC2014}. A data-dependent hypergraph partitioning algorithm \cite{zoltan2Dh_short} has potential to provide a substantial performance gain in this context. This method can speed up eigenvector computation, possibly by a significant factor, but it is an expensive procedure and its cost must be amortized in order to be effective. This is demonstrated on the left in Figure~\ref{fig:partitioning}. Implementing the eigensolver in Anasazi \cite{Anasazi2009_short}, we computed the time required to partition the graph---either randomly or using the hypergraph method---and perform a number of matrix-vector multiplications for the residuals matrix of an R-MAT graph with $2^{23}$ vertices and an average degree of 8. It requires approximately 40,000 matrix-vector multiplications to make the cost of performing the more sophisticated partitioning method worth the cost over random partitioning \cite{CSC2014}. In a dynamic setting, however, if a partition remains valid over time, it may be possible to recover the cost by reusing at several time instances, and preliminary results suggest that this may be the case. As an experiment, in the process of generating an R-MAT graph, we partitioned the vertices after only 30\% of the edges have been added. This simulates the process of partitioning a network while it is growing. As shown on the right in Figure~\ref{fig:partitioning}, a substantial gap between the running time of the matrix-vector multiplication is maintained as edges are added. When all of the edges in the graph are present, the multiplication performed on the randomly partitioned graph takes about 40\% longer than on the graph using the hypergraph partitioning method. This opens up an interesting line of research for static graphs as well: If a large graph can be sampled such that hypergraph partitioning still substantially improves performance at lower cost, it will become a more attractive option for parallel big data analysis. Future work will focus on the area of large graph sampling, both for effective partitioning in a parallel computing context and within resource-constrained data analysis environments.

\section*{Acknowledgments}
This work is sponsored by the Intelligence Advanced Research Projects Activity (IARPA) under Air Force Contract FA8721-05-C-0002. The U.S. Government is authorized to reproduce and distribute reprints for Governmental purposes notwithstanding any copyright annotation thereon.  Disclaimer: The views and conclusions contained herein are those of the author and should not be interpreted as necessarily representing the official policies or endorsements, either expressed or implied, of IARPA or the U.S. Government.

This research used resources of the National Energy Research Scientific Computing Center, which is supported by the Office of Science of the U.S. Department of Energy under Contract No. DE-AC02-05CH11231.

\bibliographystyle{plain} 
\bibliography{FullBib}

\end{document}